\documentclass{aa}  
\usepackage{graphicx}
\usepackage{txfonts}
\usepackage{lipsum}
\usepackage{subcaption}
\usepackage{lscape}
\usepackage{placeins}
\usepackage[colorlinks=true, citecolor=blue, linkcolor=black, urlcolor=black]{hyperref}
\usepackage{orcidlink}
\usepackage{bm}
\usepackage{enumitem}
\usepackage{xcolor}
\usepackage{float}

\begin{document}
\title{3D radiative transfer of K-series lines of He-like ions with \texttt{SKIRT}}

\author{
N. Sameshima\inst{1,2}\email{naoto@ac.jaxa.jp}$^{\orcidlink{0009-0005-6772-5529}}$\fnmsep\thanks{naoto@ac.jaxa.jp}
\and 
A. Lauwers\inst{3}\email{arno.lauwers@ugent.be}$^{\orcidlink{0009-0003-9692-9382}}$
\and 
B. Vander Meulen\inst{3,4,5}\email{bert.vandermeulen@esa.int}$^{\orcidlink{0000-0002-5488-1961}}$
\and 
\\
M. Tsujimoto\inst{1,2}\email{tsujimoto.masahiro@jaxa.jp}$^{\orcidlink{0000-0002-9184-5556}}$\fnmsep\thanks{tsujimoto.masahiro@jaxa.jp}
\and
L.-Y. Gu\inst{5}\email{l.gu@sron.nl}$^{\orcidlink{0000-0001-9911-7038}}$
\and 
M. Baes\inst{3}\email{maarten.baes@ugent.be}$^{\orcidlink{0000-0002-3930-2757}}$
\and 
P.~Camps\inst{3}\email{peter.camps@me.com}$^{\orcidlink{0000-0002-4479-4119}}$
\and
D.~Miura\inst{1,6}\email{dikmur0611@g.ecc.u-tokyo.ac.jp}$^{\orcidlink{0009-0009-0439-1866}}$
}
\institute{
Institute of Space and Astronautical Science, Japan Aerospace Exploration Agency, 3-1-1 Yoshino-dai, Chuo-ku, Sagamihara, Kanagawa 252-5210, Japan
\and Department of Astronomy, Graduate School of Science, The University of Tokyo, 7-3-1 Hongo, Bunkyo-ku, Tokyo 113-0033, Japan
\and Department of Physics and Astronomy, Universiteit Gent, Proeftuinstraat 86 N3, B-9000 Ghent, Belgium
\and European Space Agency, European Space Research and Technology Centre, Keplerlaan 1, 2201 AZ Noordwijk, The Netherlands
\and SRON Netherlands Institute for Space Research, Niels Bohrweg 4, 2333 CA Leiden, The
Netherlands
\and Department of Physics, Graduate School of Science, The University of Tokyo, 7-3-1 Hongo, Bunkyo-ku, Tokyo 113-0033, Japan
}
\date{\today}

\abstract
{High-resolution X-ray spectroscopy with XRISM has resolved line profiles of highly
ionised ions, including P Cygni profiles, in many systems containing compact objects. To
interpret these line profiles, forward modelling with three-dimensional radiative
transfer (3D RT) calculations including the relevant ionic processes is required.}
{Following our previous work on the Lyman-series lines of H-like ions, we present the
implementation of the K-series lines of He-like ions with $Z=2$--30 in the open-source
Monte Carlo RT code \texttt{SKIRT}, providing a numerical tool for interpreting
high-resolution X-ray spectra of photoionised plasmas.}
{Building upon the framework developed for H-like ions, we incorporate the resonance
scattering (RS) and radiative recombination (RR) channels. We also implement the
inner-shell ionisation (II) channel of Li-like ions for the production of He-like
lines. We resolve fine-structure levels and account for level populations, frequency
redistribution, phase function, and linear polarisation through transition-specific
Mueller matrices.}
{The implementation reproduces the expected RT effects in He-like lines. The
\ion{Fe}{XXV} $w$ and $y$ lines, dominated by the RS channel, develop optically thick
line-profile distortions and P Cygni profiles in an outflowing plasma, whereas the $x$
and $z$ lines, dominated by the RR and II channels, do not exhibit such behaviour. The
line intensities from the RS and RR channels agree well with \texttt{Cloudy}. The II
channel further enhances the $z$ line, which is not yet included in
\texttt{Cloudy}. The linear polarisation calculations reproduce the analytical and
experimental angular dependence of resonance scattering on a line-by-line basis.}
{This extension to He-like lines, together with our previous implementation of H-like
lines, enables 3D RT modelling of the major X-ray lines, providing a tool
for constraining gas kinematics around compact objects observed with XRISM. The
demonstrated polarisation capability also provides a foundation for future diagnostics
with X-ray polarimeters with high spectral resolution.}
\maketitle
\nolinenumbers

\section{Introduction}\label{s1}
In systems containing compact objects such as black holes (BHs) and neutron stars (NSs),
part of the gravitational energy released by mass accretion is converted into extreme-UV
and X-ray radiation \citep{D2007}. This intense radiation in turn drives mass outflows
through radiatively driven winds \citep{P2002}. The wind velocity provides information
about the transport of matter and kinetic energy from accretion into the
surrounding environment. Revealing the gas kinematics around compact objects is therefore
crucial for understanding the circulation of matter and energy in the Universe.

High-resolution X-ray spectroscopy is a powerful tool for this purpose because the
velocity structure is imprinted on spectral line profiles. This approach has been enabled
by the advent of the X-ray Imaging and Spectroscopy Mission (XRISM;
\citealt{T2025c}), which allows line profiles to be studied in unprecedented detail.
Indeed, XRISM has revealed P Cygni profiles in H-like and He-like lines of S, Ar, Ca,
and Fe in many compact-object systems \citep{C2024a,C2025c,A2025a,T2025a,C2025,M2026},
whereas previous X-ray telescopes detected them only in systems with exceptionally high
outflow velocities \citep{B2000}.

Spectral line profiles are modified by interactions between photons and outflowing gas
in three-dimensional (3D) geometries. Interpreting the observed line profiles therefore
requires 3D radiative transfer (RT) calculations. Although one-dimensional (1D) RT
codes, such as \texttt{SPEX} \citep{K2024a}, \texttt{Cloudy} \citep{G2025b}, and
\texttt{XSTAR} \citep{K1994}, have been widely used in X-ray studies, Monte Carlo RT
(MCRT) codes offer clear advantages for treating complex 3D geometries and velocity
fields \citep{S2013a,N2019}.

\texttt{SKIRT} \citep{C2015,B2015,C2020} is a state-of-the-art MCRT code with flexible
3D geometries, radiation sources, spatial grids, detectors, and interfaces to
hydrodynamical simulations. The code is available as open-source
software\footnote{\url{https://github.com/SKIRT/SKIRT9}} and is supported by extensive
documentation and tutorials\footnote{\url{https://skirt.ugent.be}}. Modern acceleration
techniques, including peel-off, forced scattering, continuous absorption, composite
biasing, and explicit absorption \citep{B2011,B2016,B2022a}, make it well suited to
interpreting high-resolution observations at feasible computational cost. The code was
originally developed for dust radiative transfer from the sub-mm to ultraviolet
bands \citep{B2003} and has since been extended to other wavelength regimes, including
molecular and atomic excitation in the millimetre band \citep{M2023c} and several
X-ray processes involving neutral matter, free electrons, and bound electrons
\citep{M2023b,M2024}.

We recently began incorporating X-ray ionic processes into \texttt{SKIRT} to extend its
applicability to high-resolution X-ray spectroscopy. The first extension implemented the
$n \rightarrow 1$ transitions of H-like ions with $Z=1$--30 \citep{S2026}. In this
paper, we extend the framework to the corresponding transitions of He-like ions. We
follow the basic scheme established for H-like ions, but several important differences
arise from the more complex atomic structure. Whereas H-like lines are modelled using
only resonance scattering (RS) and radiative recombination (RR), He-like lines also
involve inner-shell ionisation (II), dielectronic capture (DC), and autoionisation
(AI).

\medskip

The remainder of this paper is organised as follows. Section~\ref{s2} describes the
scope, physical processes, and formalism. Section~\ref{s3} presents the implementation,
Sect.~\ref{s4} the verification tests, Sect.~\ref{s5} a laboratory plasma
application, Sect.~\ref{s6} the limitations of the present work, and
Sect.~\ref{s7} summarises the results.

\section{Physics}\label{s2}
\subsection{Scope}\label{s2-1}
Our scope is low-density photoionised plasmas around compact objects. These plasmas are
assumed to be under non-local thermodynamic equilibrium (NLTE) conditions, in which
radiative processes dominate over collisional processes. The density and optical-depth 
ranges over which this approximation holds depend on the element and transitions of interest; 
we discuss some of these limitations in Sect.~\ref{s6}. A fully self-consistent treatment of
such plasmas requires RT calculations to determine the ionisation balance, level
populations, and electron temperature consistently with the radiation field and to
synthesise X-ray spectra. In the present work, however, we focus on the RT calculation
for a prescribed plasma state, with the ionisation balance and electron temperature
provided externally (Sect.~\ref{s2-3}).

In this work, we focus on the K-series ($n\rightarrow 1$) lines of He-like ions with
$Z=2$--30. The lines are denoted He$\alpha$ ($2\rightarrow1$), He$\beta$
($3\rightarrow1$), He$\gamma$ ($4\rightarrow1$), and so on. Following our previous
implementation for H-like ions \citep{S2026}, we consider upper levels up to $n=10$ and
resolve the fine-structure levels. This choice is motivated by recent XRISM
observations, which resolve fine-structure levels \citep{M2025b} and detect transitions
up to $n=9$ (He$\theta$) in \ion{Fe}{XXV} \citep{A2025a}. We also include the linear
polarisation associated with resonance scattering, because it can modify the emergent
line profiles in anisotropic geometries.

\subsection{Atomic processes}\label{s2-2}
As a basis for the following implementation, we outline the atomic levels and processes
responsible for the formation of the He-like K-series lines. We use Fe as an
illustrative example, although the same considerations apply to all He-like ions. A
general description of the relevant atomic physics in photoionised plasmas is given by
\citet{O2006}.

\paragraph{Levels and transitions}
\begin{figure}
 \begin{center}
  \includegraphics[width=1.0\columnwidth,clip]{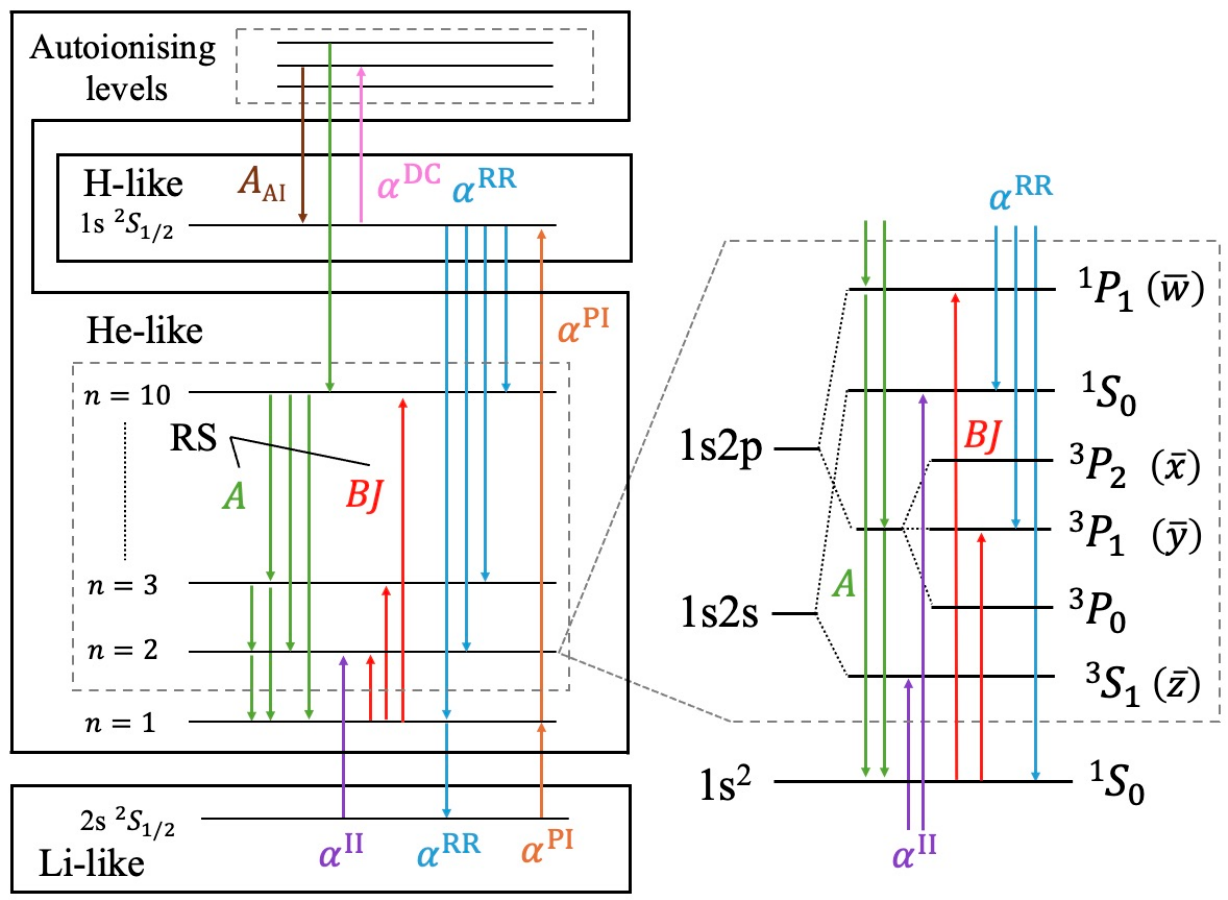}
 \end{center}
\caption{Atomic levels and processes included in the present implementation. $A$ and
$B$ denote the Einstein coefficients, $J$ is the radiation field, and $BJ$ represents
the radiative excitation rate. RS, RR, PI, II, DC, and AI denote resonance scattering,
radiative recombination, photoionisation, inner-shell ionisation, dielectronic
capture, and autoionisation, respectively. Other symbols are defined in the main text.}
\label{f01}
\end{figure}

\begin{figure}
 \begin{center}
  \includegraphics[width=1.0\columnwidth,clip]{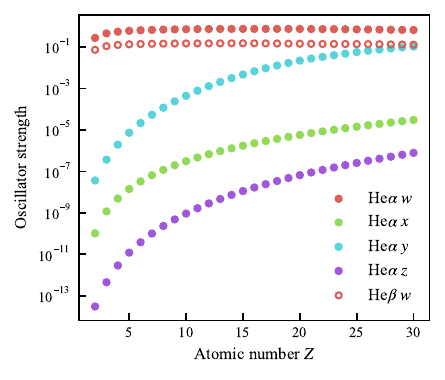}
 \end{center}
\caption{Oscillator strengths of the He$\alpha$ $w$ (red), $x$ (yellow-green), $y$
(cyan), and $z$ (purple) lines, together with the He$\beta$ $w$ line, for He-like ions
with $Z=2$--30. The atomic data are taken from the \texttt{SPEX}~v3.08.01 database
\citep{K2024a}.}
 \label{f02}
\end{figure}

\begin{figure}[H]
 \begin{center}
  \includegraphics[width=0.95\columnwidth,clip]{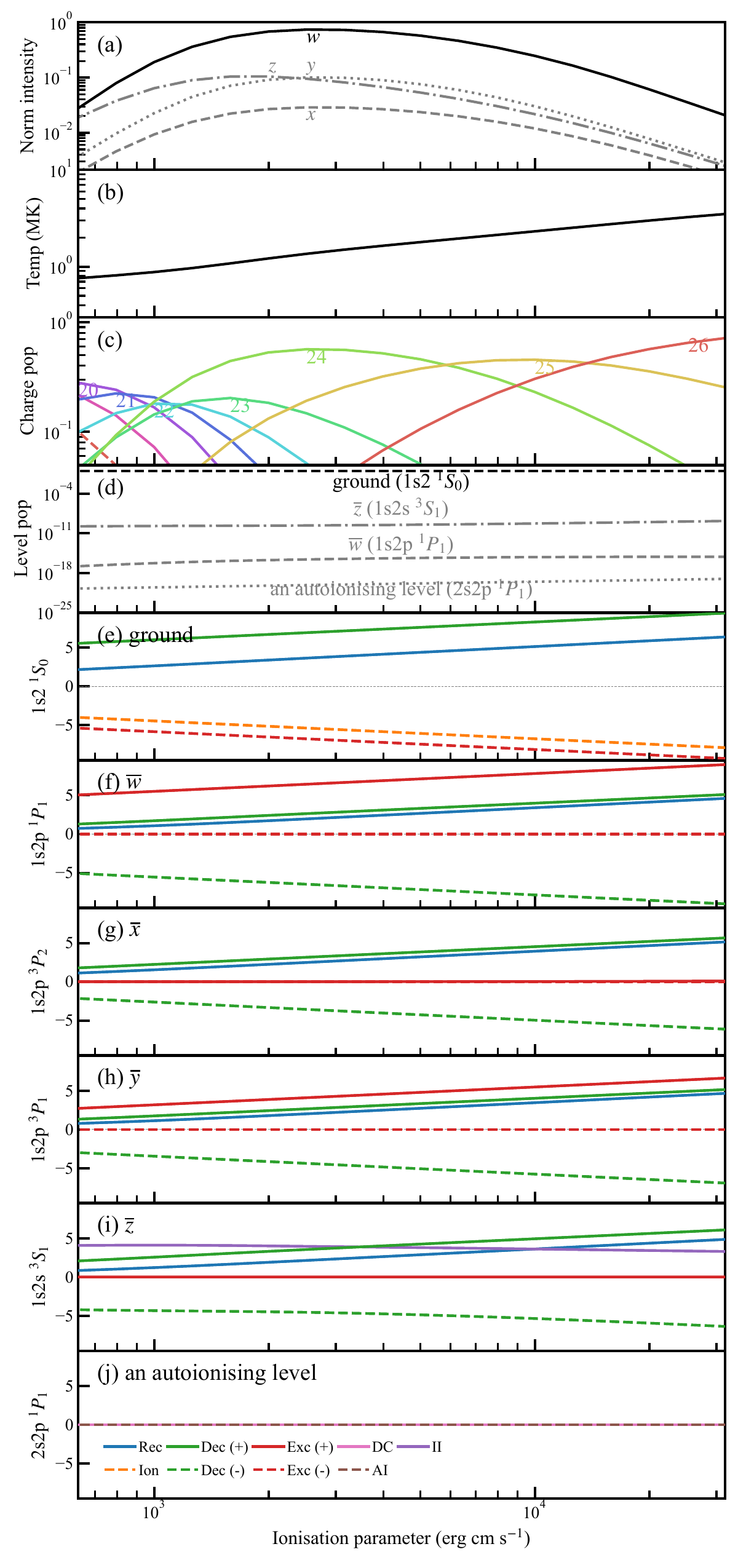}
 \end{center}
\caption{(a) Fe He$\alpha$ line intensities, (b) electron temperature, (c) charge
 populations, (d) level populations, and (e)--(j) rates of the processes that populate
 ($y>0$, solid) or depopulate ($y<0$, dashed) each level as a function of the ionisation
 parameter $\xi$. The calculations were performed with \texttt{SPEX} \citep{K2024a} in
 the low-density ($10^{10}$ cm$^{-3}$) and optically thin limit. In panel (c), the
 numbers denote the number of stripped electrons. In panels (e)--(j), the rates ($r$)
 are plotted as $y=\operatorname{arcsinh}(r)$, so that they are approximately
 logarithmic, $y\approx\log|r|$, for $|r|\gg1$ while preserving the sign (positive for
 population and negative for depopulation). The incident radiation is assumed to be a
 power-law SED with a photon index of 2 over 0.001--20~keV. The individual processes are
 described in the main text.}
 \label{f03}
\end{figure}

We first describe the relevant levels and transitions illustrated in Fig.~\ref{f01}. We
consider (i) the ground state of He-like ions ($\mathrm{1s^{2}\,^1S_0}$) together with
the ground states of the two adjacent ionisation stages, namely (ii) H-like ions
($\mathrm{1s\,^2S_{1/2}}$) and (iii) Li-like ions ($\mathrm{2s\,^2S_{1/2}}$). For the
He-like ions, we also include (iv) singly excited states up to $n=10$ and (v) doubly
excited states, in which both electrons are excited from the 1s shell.  The doubly
excited states lie above the ionisation threshold and can decay via the Auger process;
they are therefore referred to as autoionising levels. Despite this name, they can also
decay radiatively without ionisation, with radiative decay becoming dominant for
high-$Z$ ions \citep{P2003}.

The He$\alpha$ ($n=2\rightarrow1$) complex comprises the following four
transitions between fine-structure levels.
\begin{align*}
 w &: \; \mathrm{1s\,2p\,}^1P_1 \rightarrow \mathrm{1s^2\,}^1S_0 \\
 x &: \; \mathrm{1s\,2p\,}^3P_2 \rightarrow \mathrm{1s^2\,}^1S_0 \\
 y &: \; \mathrm{1s\,2p\,}^3P_1 \rightarrow \mathrm{1s^2\,}^1S_0 \\
 z &: \; \mathrm{1s\,2s\,}^3S_1 \rightarrow \mathrm{1s^2\,}^1S_0
\end{align*}

The $w$ line is an allowed electric dipole (E1) transition because it satisfies all E1
selection rules. The $z$ and $x$ lines are forbidden transitions, whose dominant decay
channels are magnetic dipole (M1) and magnetic quadrupole (M2), respectively. The $y$
line violates only the spin-selection rule ($\Delta S=0$), whose importance decreases
with increasing $Z$ because of the growing contribution of spin--orbit coupling. For
$Z=26$ (Fe), the oscillator strength of the $y$ line is within approximately one order
of magnitude of that of the $w$ line (Fig.~\ref{f02}). The $y$ line remains an
intercombination transition, but its oscillator strength approaches that of an allowed
transition as $Z$ increases. The same classification applies to the higher ($n>2$)
series, whose components are similarly denoted as, for example, He$\beta$ $w$, $x$, $y$,
and $z$.

\paragraph{Populations}
We next review the physical conditions relevant to photoionised plasmas. These
conditions are conveniently characterised by the ionisation parameter $\xi \equiv
L_\mathrm{X}/n_\mathrm{H}r^{2}$ \citep{T1973}, where $L_\mathrm{X}$ is the incident
X-ray luminosity in the 1--1000~Ryd range, $n_\mathrm{H}$ is the local hydrogen number
density, and $r$ is the distance from the X-ray source to the
plasma. Figure~\ref{f03} (a) shows the intensities of the four \ion{Fe}{XXV} He$\alpha$
lines, while Fig.~\ref{f03} (b--d) shows the corresponding plasma properties as a
function of $\xi$. The peak emissivity of the $w$ line and the peak Fe$^{+24}$ ion
fraction both occur at $\log \xi \simeq 3.3$ (panels (a) and (c)). At this ionisation
parameter, the electron temperature determined from thermal balance is approximately
1~MK (panel (b)), so that collisional processes are less important compared to radiative
processes. The ground-state population of Fe$^{+24}$ is nearly unity, whereas the
populations of the excited levels are negligible (panel (d)).

\paragraph{Processes}
We finally describe the relevant atomic processes. Panels (e)--(j) show the rates of
radiative and Auger processes that populate (solid curves) and depopulate (dashed
curves) each level. For convenience, we denote the upper level of line $a$ by $\bar{a}$
(e.g. $\bar{w}$ for $\mathrm{1s\,2p\,}^1P_1$).

The He-like ground state is populated by radiative recombination (RR) from Fe$^{+25}$
(`Rec') and by radiative decay from excited levels of Fe$^{+24}$ (`Dec~($+$)'). These
processes also populate $\bar{w}$, $\bar{x}$, $\bar{y}$, and $\bar{z}$. Radiative
excitation from the ground state (`Exc~($+$)') is the dominant population mechanism for
$\bar{w}$ and $\bar{y}$, whereas it is almost negligible for $\bar{x}$ and $\bar{z}$,
reflecting their much smaller oscillator strengths (Fig.~\ref{f02}). These processes
are also important for H-like ions. In addition, He-like ions are populated through
inner-shell ionisation (`II') of Li-like ions (Fe$^{+23}$). When a 1s electron is
photoionised from the ground state of a Li-like ion ($\mathrm{1s^2\,2s\,}^2S_{1/2}$),
the resulting He-like ion is left in either the $\mathrm{1s\,2s\,}^3S_1$ ($\bar{z}$) or
$\mathrm{1s\,2s\,}^1S_0$ excited state. The former decays predominantly via the $z$
line, whereas the latter decays mainly through two-photon emission.  This process is
most important around $\log\xi\sim3$, where the Fe$^{+23}$ ion fraction reaches its
maximum (panels (c) and (i)). The autoionising levels are primarily populated by
dielectronic capture (`DC'), in which a free electron is captured while simultaneously
exciting a bound electron, thereby producing a doubly excited state. However, DC is
negligible under the ionisation conditions considered here (panel (j)), because the
electron temperature in a photoionised plasma is too low for efficient collisional
excitation of the inner-shell electron.

The He-like ground state is depopulated by photoionisation to Fe$^{+25}$ (`Ion') and
by radiative excitation to excited levels (`Exc~($-$)'). The excited levels are
depopulated through radiative decay to lower levels (`Dec~($-$)'). These decays occur
either directly to the ground state or through cascades involving intermediate
levels. For example, $\bar{y}$ can decay directly to the ground state by emitting a
$y$-line photon, or cascade through $\bar{z}$, emitting a UV photon
($\bar{y}\rightarrow\bar{z}$) followed by a $z$-line photon. The autoionising levels are
depopulated either by autoionisation (AI), through the ejection of an Auger electron,
or by radiative decay, although both processes are negligible under the ionisation
conditions considered here (panel (j)).

\subsection{Formalism}\label{s2-3}
We formulate the atomic processes described above following the approach adopted
for the H-like ions \citep{S2026}. The treatments of RR and RS are identical to
those for H-like ions and are briefly reviewed here for completeness. The II,
AI, and DC processes are newly implemented in this work.

Throughout the following formulation, the densities of the H-like, He-like, and Li-like
ions for the elements of interest, as well as the electron temperature, are treated as
externally prescribed input parameters. These parameters can be obtained from
photoionisation-equilibrium calculations performed with 1D RT codes
such as \texttt{SPEX} \citep{K2024a}, \texttt{Cloudy} \citep{G2025b}, and \texttt{XSTAR}
\citep{K1994}.

The excited levels, including the autoionising levels, are indexed by
$i=1,2,3,\ldots$ in ascending order of energy (e.g. $i=1$ for $\bar{z}$). The
level populations, $\bm{N}=(N_1,N_2,\ldots,N_M)^\top$, are obtained by solving
the steady-state rate equation,
\begin{align}
\label{eq:rate_tot}
\dot{\bm{N}} = \bm{\Gamma}\bm{N} + \bm{S} = \bm{0}.
\end{align}
Here, $\bm{\Gamma}$ represents transitions among the excited levels and
$\bm{S}$ is the source term.

The matrix elements are given by
\begin{align}
 \label{eq:matrix}
 \Gamma_{ij}N_{j} = \sum_{j=i+1}^{M} A_{ji}N_{j} - N_i
 \left(\sum_{k=0}^{i-1}A_{ik}+A_i^\mathrm{AI}+A_i^{2\gamma}\right),
\end{align}
where $A_{ji}$ is the Einstein $A$ coefficient for the transition from level $j$ to $i$,
and $A_i^\mathrm{AI}$ and $A_i^{2\gamma}$ are the total autoionisation rate and the
two-photon decay rate of level $i$, respectively.  The first term represents the
population rate from level $j$ to $i$ (`Dec~($+$)' in Fig.~\ref{f03}), whereas the
second term represents the depopulation rate from level $i$ to all lower levels
(`Dec~($-$)' in Fig.~\ref{f03}). The latter includes all spontaneous decay processes,
namely radiative decay (including both one- and two-photon decay) and
autoionisation. Stimulated processes are neglected. Stimulated emission is negligible
compared with spontaneous emission (Eq.~(3) of \citealt{S2026}), while stimulated
absorption, namely radiative excitation from excited levels, is also negligible under
the conditions considered here (see `Exc~($-$)' for the excited levels in
Fig.~\ref{f03}).

The ground levels of the H-like (with a density $n_{\mathrm{H\mathchar`-like}}$),
He-like ($n_{\mathrm{He\mathchar`-like}}$), and Li-like
($n_{\mathrm{Li\mathchar`-like}}$) ions are not included in $\bm{N}$.  Instead, their
contributions are incorporated through the source term $\bm{S}$. The element $S_i$
represents the net rate populating excited level $i$ from these ground levels.

\paragraph{RR}
From the H-like ground level, the RR process populates level $i$ at the rate
\begin{align}
  S^{\mathrm{RR}}_i \equiv n_{\mathrm{e}}\,n_{\mathrm{H\mathchar`-like}}
  \alpha^{\mathrm{RR}}_i(T).
\end{align}
Here, $n_{\mathrm{e}}$ is the electron density and
$\alpha^{\mathrm{RR}}_i(T)$ is the RR rate coefficient into level $i$ at
electron temperature $T$. The coefficient includes contributions from both
direct radiative recombination into the excited level and dielectronic capture
into doubly excited levels.

\paragraph{RS}
From the He-like ground level, the RS process populates level $i$ at the rate
\begin{align}
  S^{\mathrm{RS}}_i \equiv
  n_{\mathrm{He\mathchar`-like}}
  B_{0i}J(\nu_{i0}),
\end{align}
where $B_{0i}$ is the Einstein $B$ coefficient for absorption from the ground
level to level $i$, and $J(\nu_{i0})$ is the mean intensity of the local
radiation field at the transition frequency $\nu_{i0}$.

\paragraph{II}
From the Li-like ground level, the II process populates two excited levels,
$\mathrm{1s\,2s\,}^3S_1$ ($i=1$ or $\bar{z}$) and $\mathrm{1s\,2s\,}^1S_0$ ($i=5$). The
total rate is determined by the inner-shell photoionisation cross section
$\alpha^{\mathrm{II}}(\nu)$ multiplied by the ionising photon density and integrated
over frequencies above the ionisation threshold $\nu_{\mathrm{th}}$. The rate is
partitioned between the two levels according to their statistical weights ($g_1=3$ and
$g_5=1$),
\begin{align}
\label{eq:source_ii}
S^{\mathrm{II}}_i=
\begin{cases}
\displaystyle
\frac{g_i}{g_1+g_5}
n_{\mathrm{Li\mathchar`-like}}
\int_{\nu_{\mathrm{th}}}^{\infty}
\frac{4\pi J(\nu)}{h\nu}
\alpha^{\mathrm{II}}(\nu)\,\mathrm{d}\nu,
&(i=1,5),\\[1.0em]
\displaystyle
0,
&(i\neq1,5).
\end{cases}
\end{align}

We treat these three source terms independently and obtain the corresponding
partial level populations by solving
\begin{align}
\bm{\Gamma}\bm{N}_{\mathrm{RS}}+\bm{S}^{\mathrm{RS}}&=\bm{0},
\label{eq:rate_rs}\\
\bm{\Gamma}\bm{N}_{\mathrm{RR}}+\bm{S}^{\mathrm{RR}}&=\bm{0},
\label{eq:rate_rr}\\
\bm{\Gamma}\bm{N}_{\mathrm{II}}+\bm{S}^{\mathrm{II}}&=\bm{0}.
\label{eq:rate_ii}
\end{align}
The solution of Eq.~(\ref{eq:rate_tot}) is then
\begin{align}
  \bm{N}=\bm{N}_{\mathrm{RS}}+\bm{N}_{\mathrm{RR}}+\bm{N}_{\mathrm{II}},
\end{align}
by linearity.

\section{Implementation}\label{s3}
We implement the atomic processes described in Sect.~\ref{s2} in \texttt{SKIRT}.
Continuum RT processes in partially ionised media, which are not the focus of this paper,
are also implemented in \texttt{SKIRT}, including photoabsorption, fluorescence, 
and Compton scattering by free and bound electrons (Lauwers et al. in prep.), as an extension 
of the previous implementation for neutral matter \citep{M2023b}. Below, we describe the 
implementation of the atomic processes responsible for the production of the He-like lines.

An interaction between an incident photon and matter is classified as either scattering
or absorption. A scattering event produces a new photon, generally with a different
frequency and propagation direction. An absorption event either destroys the photon or
is followed by the re-emission of one or more photons. In the present implementation, we
follow only photons scattered or re-emitted in the $n\rightarrow1$ transitions of
He-like ions. Photons emitted through lower-energy transitions are discarded, as they
cannot subsequently produce $n\rightarrow1$ photons. We further assume that radiative
recombination occurs immediately following photoionisation. Consequently, both genuine
scattering events and absorption followed by re-emission are treated computationally as
scattering events.

A photon packet is described by the Stokes vector\footnote{We neglect the Stokes
parameter $V$, because circular polarisation cannot be measured with current X-ray
polarimeters.}  $\bm{I}(\nu,\hat{\bm{k}})
=(I(\nu,\hat{\bm{k}}),Q(\nu,\hat{\bm{k}}),U(\nu,\hat{\bm{k}}))^{\mathrm T}$, where $\nu$
is the photon frequency and $\hat{\bm{k}}$ is the unit vector in the propagation
direction. The photon-packet properties are tracked at floating-point precision
during transport and are binned only when recorded by an instrument.
During an interaction, an incident photon packet with Stokes vector
$\bm{I}(\nu,\hat{\bm{k}})$ is converted into an emergent photon packet with Stokes
vector $\bm{I}(\nu',\hat{\bm{k}}')$. For the processes considered here, this conversion
is characterised by three quantities: the amplitude, branching ratio, and
redistribution. The atomic data adopted in this work are summarised in 
Table~\ref{tab:atomic_data}.

\begin{table*}[tbp]
  \centering
  \caption{Atomic data adopted in this work.}
  \label{tab:atomic_data}
  \begin{tabular}{ll}
    \hline
    Atomic data & Source \\
    \hline
    Transition frequencies $\nu_{i0}$
      & \texttt{SPEX} v3.08.01 \citep{K2024a} \\
    Einstein coefficients $A_{ji}$
      & \texttt{SPEX} v3.08.01 \citep{K2024a} \\
    Autoionisation rates $A_i^{\mathrm{AI}}$
      & \texttt{SPEX} v3.08.01 \citep{K2024a} \\
    Two-photon decay rates $A_i^{2\gamma}$
      & \texttt{SPEX} v3.08.01 \citep{K2024a} \\
    Photoionisation cross sections
      & \citet{V1995} \\
    Level-resolved radiative recombination coefficients
    $\alpha_i^{\mathrm{RR}}(T)$
      & \citet{M2016a}, based on \citet{B2006} \\
    Dielectronic capture rates
      & Derived through detailed balance \citep{M1999} \\
    \hline
  \end{tabular}
\end{table*}

\paragraph{Amplitude}
The amplitude determines the probability for any of the new interactions and is given by the
product of the local absorber density and the absorption cross section. For RS, we adopt
the line cross section including natural and thermal broadening represented by a Voigt
profile, following \citet{S2026}.  For RR and II, we adopt the parameterised
photoionisation cross sections\footnote{The values of the fitting parameters are available at
\url{https://www.pa.uky.edu/~verner/photo.html}.} of \citet{V1995}.

\paragraph{Branching ratio}
The branching ratio gives the probability that a given interaction produces a particular
emission line. It is defined as the emission rate of that line divided by the
corresponding absorption rate. The emission rates are determined from the upper-level
populations obtained by solving the rate equations: Eq.~(\ref{eq:rate_rs}) for RS,
Eq.~(\ref{eq:rate_rr}) for RR, and Eq.~(\ref{eq:rate_ii}) for II. The level-resolved
radiative recombination coefficient $\alpha^{\mathrm{RR}}_i(T)$, averaged over a
Maxwellian electron energy distribution at temperature $T$, is taken from the
parameterisations\footnote{The values of the fitting parameters are available at
\url{https://vizier.cds.unistra.fr/viz-bin/VizieR?-source=J/A+A/587/A84}.}  of
\citet{M2016a}, which are based on the Atomic Data and Analysis Structure (ADAS)
database \citep{B2006}. The dielectronic capture rate is calculated from the
autoionisation rate using the principle of detailed balance \citep{M1999}.

\paragraph{Redistribution}
The redistribution determines the frequency, propagation direction, and
polarisation state of the emergent photon relative to the incident photon.
For RS, we follow the formalism of \citet{H1962}. If the incident (absorbed)
and emergent (re-emitted) photons correspond to the same bound--bound
transition, the scattering is coherent and the emergent photon retains the
memory of the incident photon. The emergent photon therefore retains the 
incident photon frequency in the atomic rest frame, while its frequency in 
the observer's frame is Doppler-shifted according to the atom's velocity, 
sampled following \citet{C2021}. 
This redistribution formalism naturally accounts for the behaviour in both
the line core and wings without requiring separate approximate treatments. 
The angular redistribution and the associated linear
polarisation are described by a linear combination of the monopole and
dipole Mueller matrices ($\bm{M}^{(m)}$ and $\bm{M}^{(d)}$) as
\begin{eqnarray}
 \label{eq:muller}
  \bm{M}(\theta) &=& E_1 \bm{M}^{(m)} (\theta)
                  +E_2 \bm{M}^{(d)} (\theta) \nonumber \\
              &=& E_1 \frac{1}{2}
\left(
\begin{matrix}
1&0&0\\
0&0&0\\
0&0&0
\end{matrix}
\right)\\
            &+& E_2 \frac{3}{8}
\left(
\begin{matrix}
1+\cos^2\theta & -\sin^2\theta & 0\\
-\sin^2\theta & 1+\cos^2\theta & 0\\
0 & 0 & 2\cos\theta
\end{matrix}
\right)\nonumber .
\end{eqnarray}
Here, $\theta$ is the angle between the incident and emergent photon
directions. \citet{H1947} derived the coefficients $E_1$ and $E_2$ ($E_1+E_2=1$) as
functions of the total angular momenta of the lower and upper levels ($J$ and $J'$) and
their difference ($\Delta J\equiv J'-J$), as listed in Table~\ref{t01}. We apply this
formalism to all transitions with $|\Delta J|\le1$, since RS is important only for
E1 transitions, for which $\Delta J=\{-1,0,1\}$. For the much weaker
transitions with $|\Delta J|\ge2$, we adopt isotropic (monopole) scattering ($E_1=1$,
$E_2=0$).

Our RS implementation also treats the case in which the incident and emergent photons
correspond to different bound--bound transitions. In this case, the frequency redistribution 
is incoherent in the atomic rest frame, with the emergent photon frequency in that frame set 
by the emission transition frequency. The scattering-atom velocity is sampled as in the 
coherent case and determines the Doppler shift of the emergent photon in the observer's frame. 
The angular distribution is assumed to be purely
monopole. The same treatment is adopted for the RR and II channels.

\begin{table}
 \centering
 \renewcommand{\arraystretch}{2.3}
 \caption{Weights of the monopole ($E_1$) and dipole components ($E_2$) in the
 Mueller matrices.}
    \label{t01}
    \begin{tabular}{ccc}
        \hline
        $\Delta J$ & $E_1$ & $E_2$ \\
        \hline
        $1$ &
        $\displaystyle \frac{1}{10}\frac{3J(6J+7)}{(J+1)(2J+1)}$ &
        $\displaystyle \frac{1}{10}\frac{(2J+5)(J+2)}{(J+1)(2J+1)}$ \\[1ex]
        $0$ &
        $\displaystyle \frac{1}{10}\frac{3(2J^2+2J+1)}{J(J+1)}$ &
        $\displaystyle \frac{1}{10}\frac{(2J-1)(2J+3)}{J(J+1)}$ \\[1ex]
        $-1$ &
        $\displaystyle \frac{1}{10}\frac{3(J+1)(6J-1)}{J(2J+1)}$ &
        $\displaystyle \frac{1}{10}\frac{(2J-3)(J-1)}{J(2J+1)}$ \\[1ex]
        \hline
    \end{tabular}
    \tablefoot{ The values in \cite{H1947} are recalculated to be applicable to
 Eq.~(\ref{eq:muller}). $J$ denotes the total angular momentum of the lower level and
 $\Delta J$ denotes the difference in total angular momentum between the lower and
 upper levels of the transition.}
\end{table}

\section{Verification}\label{s4}
We now validate the implementation described in Sect.~\ref{s3}. We compare synthetic
\texttt{SKIRT} spectra with analytical predictions and with the results of a 1D RT
code. Because the implementation relies on \texttt{SPEX} \citep{K2024a}, we use
\texttt{Cloudy} c25 \citep{G2025b} as an independent benchmark.

The four He$\alpha$ lines ($w$, $x$, $y$, and $z$) have markedly different oscillator
strengths (Fig.~\ref{f02}), and consequently the dominant population and depopulation
mechanisms of their upper levels differ (Fig.~\ref{f03}). These differences give rise to
distinct RT effects on the line intensities and profiles, providing powerful diagnostics
of photoionised plasmas when modelled properly. We demonstrate two such effects: the
dependence of the line intensities on optical depth (Sect.~\ref{s4-1}) and the
modification of the line profiles by a velocity field (Sect.~\ref{s4-2}) using the
\ion{Fe}{XXV} lines.

In both cases, we consider a spherical shell of photoionised gas consisting of 
H-like, He-like, and Li-like Fe ions illuminated by an
isotropic, unpolarised central point source with a flat spectral energy distribution
(SED). In the \texttt{SKIRT} simulations, we used $10^{10}$ photon packets,
100 radial grid cells, and $0.1$~eV output spectral bins for each run. The H-like, He-like, and Li-like ion densities,
$n_{\mathrm{H\mathchar`-like}}$, $n_{\mathrm{He\mathchar`-like}}$, and
$n_{\mathrm{Li\mathchar`-like}}$, together with the electron temperature $T$, were
computed with \texttt{Cloudy} and imported into each radial cell of the \texttt{SKIRT}
model. The populations of the excited levels of the He-like ions were then computed by
solving the rate equations (Eqs.~(\ref{eq:rate_rs})--(\ref{eq:rate_ii})). In the
\texttt{Cloudy} calculations, all of these quantities were computed self-consistently.

\subsection{Optical depth effects}\label{s4-1}
\begin{figure}[ht]
 \begin{center}
  \includegraphics[width=0.9\columnwidth,clip]{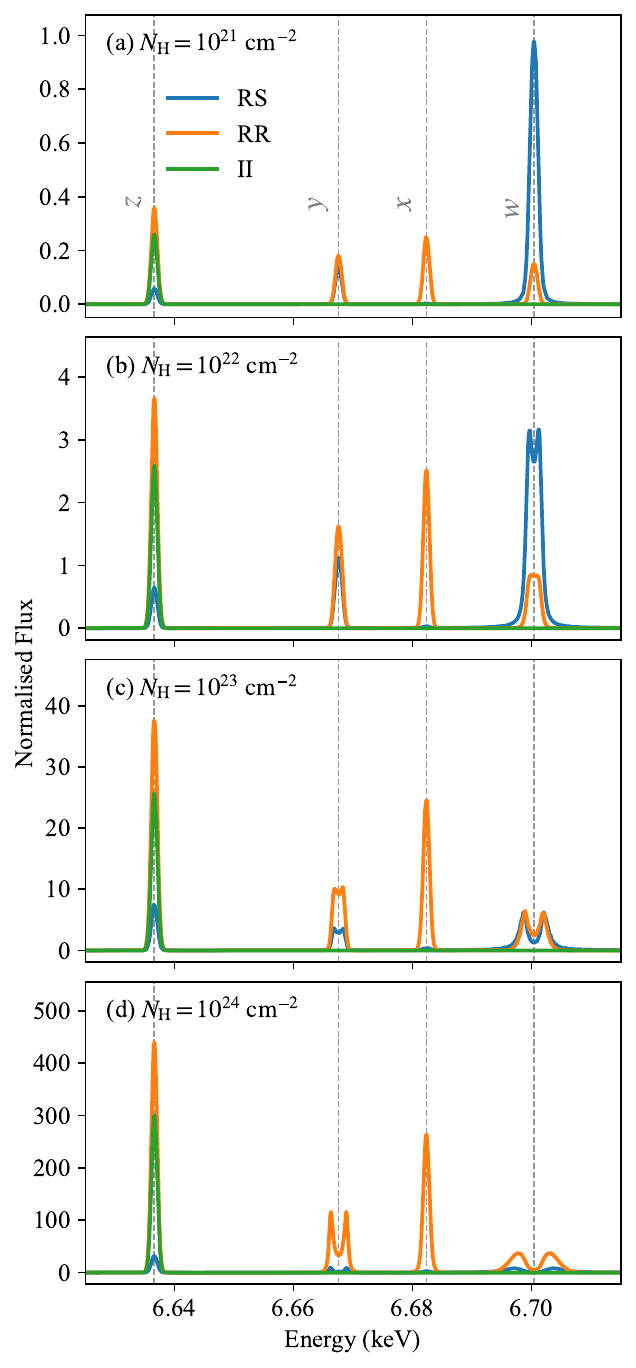}
 \end{center}
 \caption{Fe He$\alpha$ line profiles calculated with \texttt{SKIRT} for column densities
 of $10^{21}$, $10^{22}$, $10^{23}$, and $10^{24}\,\mathrm{cm^{-2}}$, shown in panels
 (a)--(d), respectively. The contributions from resonance scattering (RS; blue),
 radiative recombination (RR; orange), and inner-shell ionisation (II; green), classified
 according to the channel of the initial interaction, are shown separately. The
 intensities are normalised to the RS contribution to the $w$ line at
 $N_\mathrm{H}=10^{21}\,\mathrm{cm^{-2}}$.}
 \label{f04}
\end{figure}

\begin{figure}[ht]
 \begin{center}
  \includegraphics[width=1.0\columnwidth,clip]{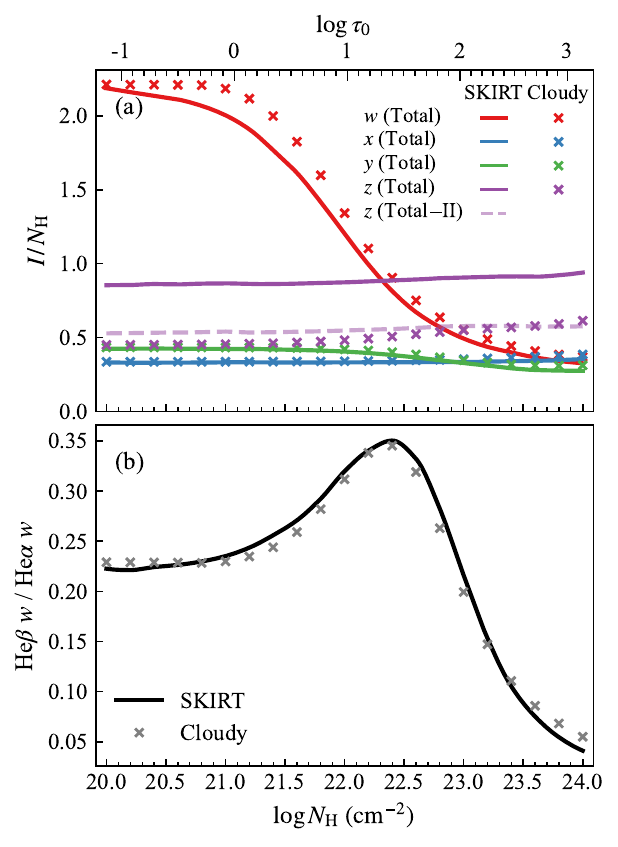}
 \end{center}
\caption{(a) Fe He$\alpha$ $w$ (red), $x$ (blue), $y$ (green), and $z$ (purple) line
intensities ($I$) normalised by the hydrogen column density ($N_{\mathrm{H}}$),
calculated with \texttt{SKIRT} (solid curves) and \texttt{Cloudy} (crosses). The dashed
curve shows the \texttt{SKIRT} prediction for the $z$ line with the contribution from
inner-shell ionisation (II) excluded (`Total$-$II'). (b) Fe He$\beta$~$w$/He$\alpha$~$w$
line ratio calculated with \texttt{SKIRT} (black curve) and \texttt{Cloudy} (grey
crosses). The lower horizontal axis gives the hydrogen column density
($N_{\mathrm{H}}$), while the upper axis shows the corresponding line-centre optical
depth of the He$\alpha$~$w$ transition ($\tau_0$).}
 \label{f05}
\end{figure}

\subsubsection{Line profile distortion}
Photons at the resonance-line frequencies experience large optical depths and become
effectively trapped in the gas through numerous resonance-scattering events. They escape
only after repeated frequency redistribution shifts them sufficiently far from the line
centre. As a result, the emergent line profile is strongly distorted, with a suppressed
core and enhanced wings. Such profiles have been derived analytically
\citep{N1990,H1973,D2006} and reproduced by 3D MCRT calculations for H-like ions with
\texttt{SKIRT} \citep{C2021,S2026}. Most 1D RT codes, however, adopt the escape
probability approximation, in which frequency redistribution by resonance scattering is
not treated explicitly.

We examine this effect using the present implementation for He-like ions. We consider a
spherical shell with a uniform density of $10^{10}~\mathrm{cm}^{-3}$ and vary the
hydrogen column density over
$N_{\mathrm{H}}=10^{20}$--$10^{24}~\mathrm{cm}^{-2}$ in steps of 0.2 dex. The
ionisation parameter is fixed at $\xi=10^{3.1}$~erg~s$^{-1}$~cm, where all three
production channels contribute significantly (Fig.~\ref{f03}).

The synthetic emission spectra in the Fe He$\alpha$ band are shown in
Fig.~\ref{f04}, with the contributions from the three production channels displayed
separately. For the $w$ and $y$ lines, the dominant production mechanisms are RS and RR
(Fig.~\ref{f03}, panels (f) and (h)). Their line profiles are distorted by repeated
resonance scattering. This effect becomes apparent for the $w$ line at
$N_\mathrm{H}\gtrsim10^{22}~\mathrm{cm}^{-2}$ and for the $y$ line at
$N_\mathrm{H}\gtrsim10^{23}~\mathrm{cm}^{-2}$, reflecting the order-of-magnitude 
difference in their oscillator strengths (Fig.~\ref{f02}). By contrast,
resonance scattering is negligible for the $x$ and $z$ lines because their oscillator
strengths are many orders of magnitude smaller. Consequently, their line profiles remain
essentially unchanged over the investigated range of $N_\mathrm{H}$.

\subsubsection{Line intensity}
We next evaluated the integrated line intensity ($I$) as a function of
$N_\mathrm{H}$ and compared the results with those from \texttt{Cloudy}. Since, to
first order, $I$ is proportional to the amount of emitting gas, we plot
$I/N_\mathrm{H}$ as a function of $N_\mathrm{H}$ to highlight deviations from linearity
over a wide dynamic range.

The results of the two codes are compared in Fig.~\ref{f05} (a). They show broadly
similar trends. For the $w$ line, and to a lesser extent the $y$ line,
$I/N_\mathrm{H}$ decreases with increasing column density. This behaviour arises because
the absorption at the resonance-line frequencies becomes saturated, so the number of
absorbed photons no longer increases linearly with column density. Radiative cascades
(e.g. $\bar{y}\rightarrow\bar{z}\rightarrow$ ground) further reduce the emitted intensity
of these lines.

For the $x$ and $z$ lines, $I/N_\mathrm{H}$ remains nearly constant with column density,
because they are produced mainly through the RR and II channels. These channels remain
optically thin even at
$N_\mathrm{H}=10^{24}~\mathrm{cm}^{-2}$, so their production rates remain
approximately proportional to $N_\mathrm{H}$.

There are nevertheless several noticeable differences between the two codes. First,
the predicted $I/N_\mathrm{H}$ of the $z$ line differs significantly. This is a known
limitation of \texttt{Cloudy}, which does not yet include the II channel for producing
the $z$ line \citep{M2025a}. Indeed, the \texttt{SKIRT} calculation performed without
the II channel (dashed curve in Fig.~\ref{f05} (a)) agrees much better with the
\texttt{Cloudy} result.

In the optically thin limit, the two codes agree well for the $x$ and $y$ lines, but
less well for the $w$ and $z$ lines. Such differences may arise either from the adopted
interaction amplitudes or from the branching ratios. We verified the interaction
amplitudes by repeating the tests described in Sect.~3 of \citet{S2026} and found
excellent agreement between the two codes for all channels. We therefore attribute the
remaining differences primarily to the branching ratios, which may reflect differences
in the level-population calculations and atomic data adopted by the two codes.

\subsubsection{Line ratio}
Finally, we examined the He$\beta$~$w$/He$\alpha$~$w$ ratio, which has been proposed as
a diagnostic of column density \citep{C2020a}. The ratio initially increases with
column density because the larger optical depth of He$\alpha$~$w$ causes its intensity
to saturate earlier than that of He$\beta$~$w$. At still higher column densities,
however, repeated resonance scattering degrades He$\beta$~$w$ photons into lower-energy
photons, causing the ratio to decline towards zero. The \texttt{SKIRT} calculation
reproduces this behaviour and agrees well with the \texttt{Cloudy} result
(Fig.~\ref{f05} (b)).

\subsection{Velocity effects}\label{s4-2}
\begin{figure}
 \begin{center}
  \includegraphics[width=1.0\columnwidth,clip]{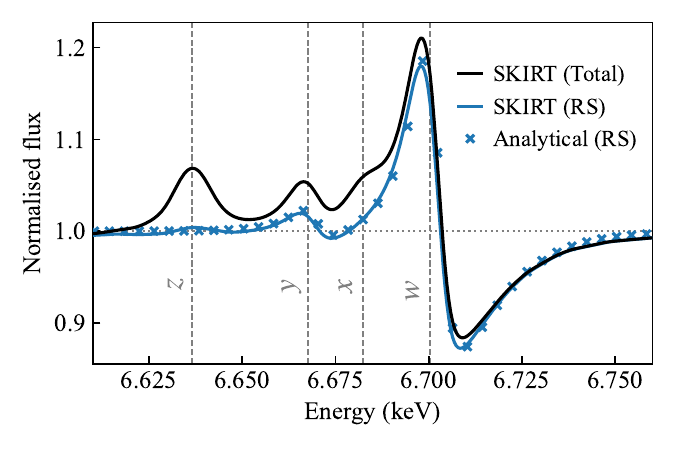}
 \end{center}
\caption{Fe He$\alpha$ line profiles in a homologously expanding spherical flow,
calculated with \texttt{SKIRT} (black curve) and compared with the analytical solution
of \citet{J1990} (blue crosses). The \texttt{SKIRT} spectrum including only the
resonance-scattering (RS) contribution is also shown (blue curve). For display purposes,
all spectra are convolved with a Gaussian of 5~eV FWHM and rebinned to 0.5~eV,
comparable to the spectral resolution of XRISM. The horizontal dotted line indicates the
incident continuum level, and the vertical dashed lines mark the rest-frame line
energies. The model parameters are: time since explosion, 3000~s; maximum ejecta
velocity, $0.01c$; photospheric velocity, $5\times10^{-4}c$; line-centre optical depth,
0.2; and density-stratification parameter, $5.0\times10^{7}$~cm~s$^{-1}$.}
\label{f06}
\end{figure}

In outflowing gas, resonance lines often exhibit P Cygni profiles, consisting of
blue-shifted absorption produced by gas moving towards the observer in the transmitted
spectrum together with emission centred on the rest wavelength in the scattered
spectrum. For the \ion{Fe}{XXV} He$\alpha$ complex, this behaviour is expected for the
$w$ and $y$ lines through the RS channel, but not for the $x$ and $z$ lines because
their resonance scattering is negligible and they therefore do not produce absorption
features.

As a benchmark, we calculated analytical P Cygni profiles using a publicly available
tool originally developed for supernova ejecta.\footnote{The code is available at
\url{https://github.com/unoebauer/public-astro-tools/tree/master/pcygni_profile}. An
example application is given by \citet{N2019}.} The calculation assumes a homologously
expanding spherical flow and evaluates the line profiles under the Sobolev
approximation, which is appropriate for flows with large velocity gradients
\citep{J1990}. The optical depth is prescribed as an exponential function of position,
following \citet{T2011}.

We performed \texttt{SKIRT} simulations using the same velocity and density
distributions as in the analytical calculation. Because the analytical code computes
only resonance-line profiles, the comparison was restricted to the $w$ and $y$ lines.
The \texttt{SKIRT} spectra from the RS channel show excellent agreement with the
analytical profiles for both lines (Fig.~\ref{f06}). The total spectra exhibit
additional emission in all four lines owing to the RR and II channels. As expected, the
$x$ and $z$ lines do not develop P Cygni profiles.

\section{Demonstration}\label{s5}
\begin{figure}
 \begin{center}
  \includegraphics[width=1.0\columnwidth,clip]{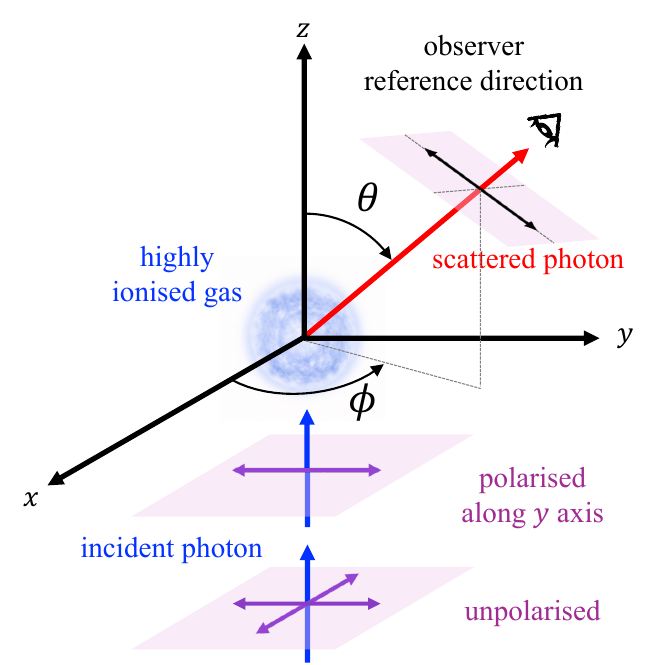}
 \end{center}
 \caption{Schematic view of the \texttt{SKIRT} simulation motivated by the experiments
 of \cite{R2013}.}
\label{f07}
\end{figure}

\begin{figure*}
 \begin{center}
  \includegraphics[width=2.0\columnwidth,clip]{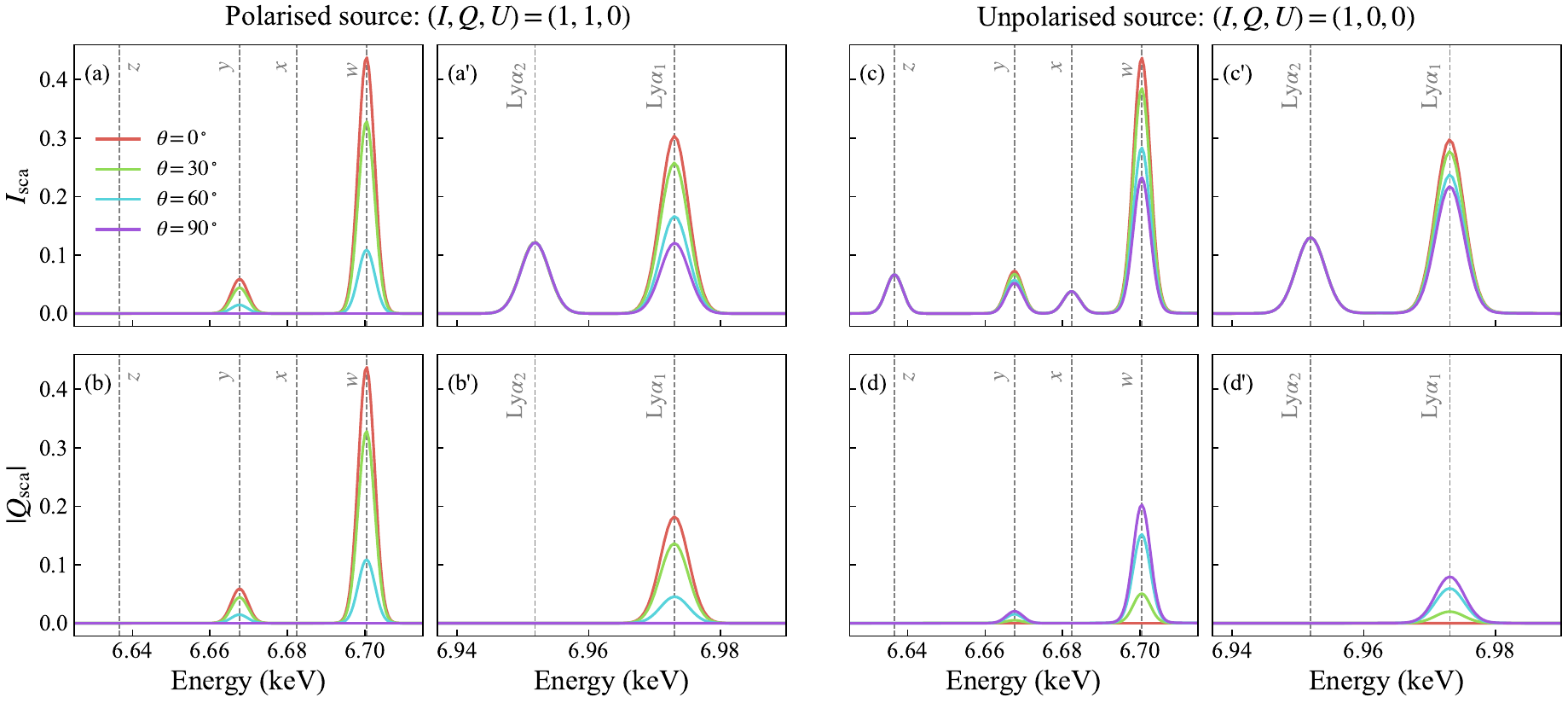}
 \end{center}
 \caption{Synthesised scattered spectra in the Fe He$\alpha$ and Ly$\alpha$ bands in the
 $I_\mathrm{sca}$ and $|Q_\mathrm{sca}|$ Stokes parameters for (a, b) a completely
 linearly polarised incident source and (c, d) a completely unpolarised incident
 source. The spectra are shown for $\phi=\pi/2$ and $\theta=0^\circ$ (red), $30^\circ$
 (green), $60^\circ$ (cyan), and $90^\circ$ (purple). The observer's reference direction
 is chosen to lie in the $y$--$z$ plane. For display purposes, the spectra are convolved
 with a Gaussian of 5~eV FWHM and rebinned to 0.5~eV, comparable to the spectral
 resolution of XRISM. The vertical dashed lines mark the rest-frame line energies.}
 \label{f08}
\end{figure*}

\begin{figure*}
 \begin{center}
  \includegraphics[width=2.0\columnwidth,clip]{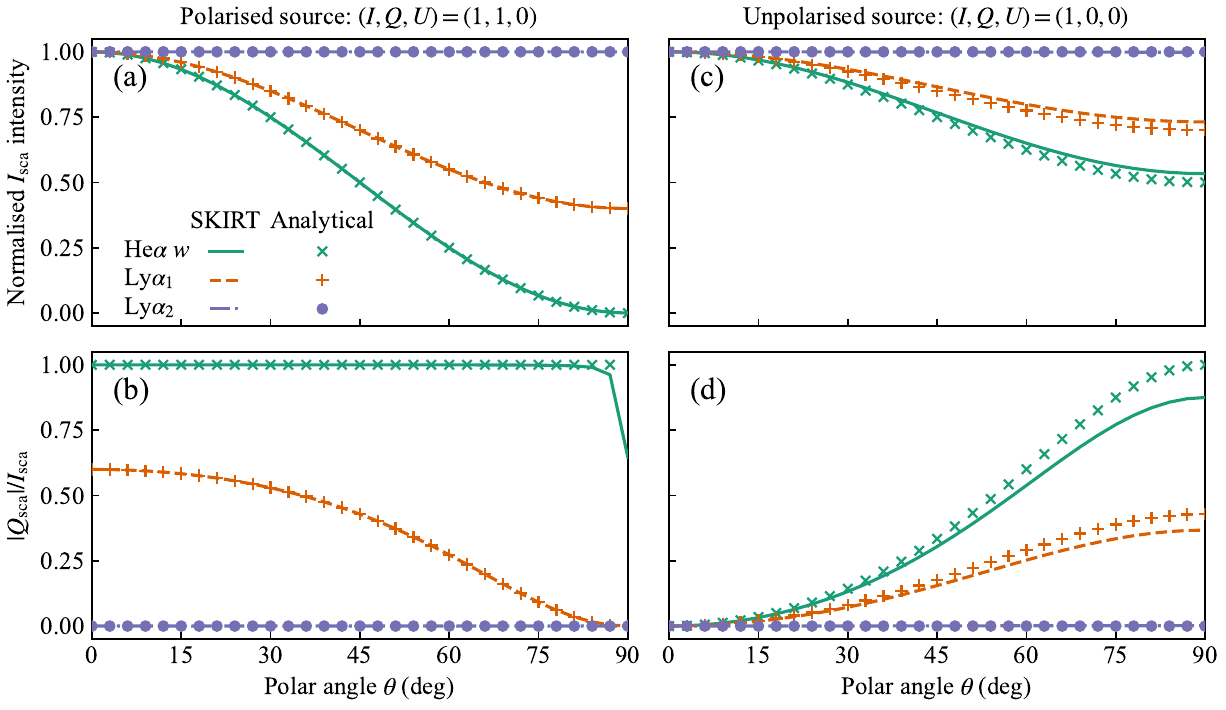}
 \end{center}
 \caption{Normalised scattered intensities ($I_\mathrm{sca}$) and polarisation degrees
 ($|Q_\mathrm{sca}|/I_\mathrm{sca}$) as a function of the polar angle ($\theta$) for a
 fixed azimuthal angle of $\phi=\pi/2$, shown in the same panel layout as
 Fig.~\ref{f08}. The \texttt{SKIRT} results (curves) are compared with the analytical
 predictions of single scattering based on Eq.~(\ref{eq:muller}) (points) for the
 He$\alpha$~$w$, Ly$\alpha_1$, and Ly$\alpha_2$ lines.}
 \label{f09}
\end{figure*}

\subsection{Setup}\label{s5-1}
We present a demonstration of the linear polarisation induced by
resonance scattering. We adopt a simple setup motivated by the laboratory experiment 
of \citet{R2013}, providing a clear illustration of how the properties of the resonance 
lines depend on the scattering direction. Although the geometry considered 
here is simple, the same calculations can be applied to more complex models with
3D geometries and velocity structures.

In this experiment, highly charged Fe ions from He-like to F-like charge states
(Fe$^{+24}$--Fe$^{+17}$) were produced and trapped in an electron beam ion trap (EBIT),
which is placed in a beam line of the PETRA III synchrotron facility. The ions are
resonantly excited by the linearly polarised synchrotron X-ray emission. The injected
X-ray energy was monochromatised using a double-crystal monochromator, which was scanned
across an energy range to cover their $n=2 \rightarrow 1$ transitions (Fig. 1 in
\citealt{R2013}). Two Ge detectors are placed perpendicular to the incident X-ray
emission; one is along the polarisation plane and the other perpendicular to it. The
ratio of the line intensity observed by the two detectors is different line by line,
reflecting their difference in the $E_1$ and $E_2$ coefficients in
Eq.~(\ref{eq:muller}). Our \texttt{SKIRT} implementation should be able to explain the
result.

\subsection{Simulation}\label{s5-2}
We performed \texttt{SKIRT} simulations using the setup shown in Fig.~\ref{f07}. A
highly ionised spherical gas consisting of Fe$^{+24}$ and Fe$^{+25}$ with equal
abundances was placed at the origin. The gas has a uniform density and a temperature of
$10^{6}$~K. Its column density through the centre corresponds to an optical depth of
0.001 in the He$\alpha$ $w$ line. The incident source is a parallel photon beam
propagating in the positive $z$ direction.

We considered two cases for the incident radiation: one representing the laboratory
experiment and the other a putative photoionising source around compact objects. In the
first case, completely linearly polarised photons are injected with the electric vector
parallel to the $y$ axis, corresponding to Stokes parameters
$(I,Q,U)=(1,1,0)$ in the reference frame defined by the $y$ axis. The incident radiation
is monochromatic, and a separate \texttt{SKIRT} simulation is performed for each
incident photon energy across the energy range of interest. The resulting spectra are
then combined to produce the final synthetic spectrum. In the second case, completely
unpolarised radiation with $(I,Q,U)=(1,0,0)$ is adopted. The incident spectrum is
continuous, so the full synthetic spectrum is obtained in a single \texttt{SKIRT}
simulation.

In both cases, \texttt{SKIRT} detectors are placed at different polar ($\theta$) and
azimuthal ($\phi$) angles. In the laboratory configuration, the gas sphere at the origin
represents the EBIT plasma, the incident monochromatic and linearly polarised beam
represents the synchrotron radiation, and the two Ge detectors correspond to viewing
directions at $(\theta,\phi)=(\pi/2,0)$ along the $x$ axis and $(\pi/2,\pi/2)$ along the
$y$ axis.

Each simulation used $10^{10}$ photon packets, 100 radial grid cells, and
$0.3$~eV output spectral bins. We extracted the scattered spectra
separately for the Stokes $I$ and $Q$ parameters in the observer's frame at each viewing
direction.  We focus on the He$\alpha$ $w$ and $y$ lines implemented in the present work
together with the Ly$\alpha_1$ and Ly$\alpha_2$ lines implemented by
\citet{S2026}. Their corresponding coefficients are $(E_1,E_2)=(0,1)$ for the $w$ and
$y$ lines (pure dipole scattering), $(1,0)$ for Ly$\alpha_2$ (pure monopole scattering),
and $(0.5,0.5)$ for Ly$\alpha_1$ (equal monopole and dipole contributions).

Although Fe is used here as an illustrative example, the $E_1$ and $E_2$ coefficients
depend only on the angular momenta of the lower and upper levels. Consequently, the same
formalism applies to all H-like and He-like ions included in the present implementation.

\subsection{Results}\label{s5-3}
Fig.~\ref{f08} shows the scattered spectra synthesised with \texttt{SKIRT}. The upper
and lower panels show the Stokes $I$ and $Q$ spectra, respectively. The left-hand panels
correspond to a linearly polarised incident source, while the right-hand panels show the
results for an unpolarised incident source. Fig.~\ref{f09} presents the normalised
scattered intensity ($I$; upper panels) and polarisation degree ($|Q|/I$; lower panels)
as a function of $\theta$ for each resonance line. Since only the degree of linear
polarisation is of interest here, we use the absolute value of the Stokes $Q$ parameter.

In Figs.~\ref{f08} and \ref{f09}, panels (a) correspond to the existing laboratory
experiment of \citet{R2013}. Panels (b) illustrate a possible future laboratory
experiment equipped with a polarisation-sensitive detector. Panels (c) represent the
case observable with XRISM, while panels (d) show the corresponding results expected
from a future mission combining the spectral resolution of XRISM with the linear
polarimetry capability of IXPE \citep{W2022}.

\paragraph{Polarised incident source (panels (a) and (b))}
In the He$\alpha$ complex, the $w$ and $y$ lines are strongly polarised, reaching
$|Q_\mathrm{sca}|/I_\mathrm{sca}\simeq1$ over most scattering angles. This reflects
their purely dipole scattering nature and is consistent with the laboratory measurements
shown in Fig.~3 of \citet{R2013}. In Fig.~\ref{f09} (b), the polarisation degree
decreases around $\theta=90^\circ$. This is because the scattered intensity is nearly zero
towards this direction, thus multiple scattering becomes more important which partially
depolarises the scattered radiation.

In the Ly$\alpha$ complex, the Ly$\alpha_1$ line exhibits a smaller polarisation degree
$|Q_\mathrm{sca}|/I_\mathrm{sca}$ than the He$\alpha$ $w$ line (Fig.~\ref{f09} (b))
because its resonance scattering consists of equal dipole and monopole components. The
isotropic component is clearly visible in $I_\mathrm{sca}$ at $\theta=90^\circ$, where
the dipole contribution vanishes (Fig.~\ref{f08} (a')). By contrast, the Ly$\alpha_2$
line remains completely unpolarised ($|Q_\mathrm{sca}|=0$), reflecting its purely
monopole scattering.

The experiment of \citet{R2013} did not cover the Ly$\alpha$ energy range. However, the
Li-like $q$ ($\mathrm{2s}\,^2S_{1/2}\rightarrow\mathrm{1s\ 2s\ 2p}\,^2P_{3/2}$) and $r$
($\mathrm{2s}\,^2S_{1/2}\rightarrow\mathrm{1s\ 2s\ 2p}\,^2P_{1/2}$) lines have the same $(E_1,E_2)$
coefficients as Ly$\alpha_1$ and Ly$\alpha_2$, respectively, and exhibit the same
angular dependence predicted here. These Li-like inner-shell transitions will be
incorporated into \texttt{SKIRT} in a future paper.

\paragraph{Unpolarised incident source (panels (c) and (d))}
An unpolarised incident beam can be regarded as an incoherent superposition of two
orthogonal linearly polarised components with equal intensity. The scattered radiation
is therefore the sum of the spectra obtained for incident polarisation vectors parallel
to the $x$ and $y$ axes. Even when the detector is insensitive to polarisation
(panel (c)), the different scattering phase functions produce observable line-to-line
differences. These differences become even more pronounced with a
polarisation-sensitive detector (panel (d)).

Unlike the polarised case, the $x$ and $z$ lines are also present in the He$\alpha$
complex in Fig.~\ref{f08} (c, d). This is because the incident radiation is assumed to be
a continuum, which contains photons energetic enough to ionise He-like and Li-like ions,
thereby producing these lines through the RR and II channels. Together with the
Ly$\alpha_2$ line, the $x$ and $z$ lines undergo purely monopole scattering and
therefore show no angular dependence on $\theta$.

The scattered intensity $I_\mathrm{sca}$ of the $w$ and Ly$\alpha_1$ lines decreases
with increasing scattering angle (Fig.~\ref{f09} (c)), although less strongly than for
the polarised incident beam (Fig.~\ref{f09} (a)). This weaker angular dependence arises
because the two orthogonal incident polarisation components contribute equally,
partially averaging out the anisotropy. The contribution from the RR channel due to 
the continuous incident spectrum causes $I_\mathrm{sca}$ to slightly exceed the analytical result.

The polarisation degree exhibits the opposite angular behaviour to that of the
polarised-source case. At $\theta=0^\circ$, two incident polarisation components
orthogonal to each other are scattered equally into the line of sight, resulting in zero
net linear polarisation. At $\theta=90^\circ$, however, the component polarised parallel
to the scattering plane is suppressed, so the scattered radiation is dominated by its
orthogonal component. This produces a strong negative Stokes parameter
($Q_\mathrm{sca}<0$), corresponding to a high degree of linear polarisation.

\section{Discussion}\label{s6}
Together with our previous work \citep{S2026}, \texttt{SKIRT} now supports the K-series
($n \rightarrow 1$) lines of both H-like and He-like ions, covering the majority of
prominent X-ray emission and absorption lines observed around compact objects. This
extension enables detailed modelling of the resolved line profiles observed with XRISM
and expected with NewAthena \citep{C2025b}, providing a new tool to investigate the
three-dimensional kinematics of photoionised plasmas. We also demonstrated that the
implementation is applicable to laboratory plasma experiments. Nevertheless, the current
implementation has several important limitations. We do not provide generic quantitative 
limits of applicability here, since these depend on the specific use case. The applicability 
can be evaluated taking into account the physical processes omitted from the present 
implementation, which we describe below.

First, the code does not solve for the ionisation balance or electron temperature.
Instead, the densities of the relevant ions and the electron temperature in each spatial
cell are supplied externally, typically from 1D photoionisation codes such as
\texttt{SPEX} \citep{K2024a}, \texttt{Cloudy} \citep{G2025b}, or \texttt{XSTAR}
\citep{K1994}, and mapped onto the 3D geometry. We also assume that nearly all He-like
ions reside in the ground level. This approximation is well justified under the
conditions considered here (Fig.~\ref{f03} (d)), but the overall calculation is therefore
not fully self-consistent.

Second, the present implementation is intended only for photoionised plasmas and is
not directly applicable to plasmas in collisional ionisation equilibrium (CIE). In CIE,
level populations are governed primarily by collisional excitation and de-excitation,
and processes such as dielectronic capture and autoionisation become much more
important. Extending the framework to CIE would therefore require substantial
modifications. Nevertheless, some aspects of the implementation, such as resonance
scattering and the resulting line-profile formation, remain applicable irrespective of
the ionisation mechanism.

Third, we neglect transitions among excited levels. The relative
intensities of the $x$, $y$, and $z$ lines in the He$\alpha$ complex provide a useful
diagnostic of the local electron density \citep{G1969} or UV radiation field
\citep{P2007}, but this is not supported in the present
implementation. The intensities of these lines are modified by the collisional or
radiative excitation from $\bar{z}$ to $\bar{x}$ or $\bar{y}$. The collisional
excitation becomes important at high densities; for example, the critical density is
$6.7\times10^{8}\ {\rm cm^{-3}}$ for \ion{C}{V} and $4.7\times10^{16}\ {\rm cm^{-3}}$
for \ion{Fe}{XXV} \citep{B1972}. The radiative excitation becomes important in intense
UV radiation fields, such as those near O-type stars \citep{R2026} for low-$Z$ elements
and the neutron star surface \citep{M2025a} for high-$Z$ elements including Fe. At line optical depths
higher than those verified in Sect.~\ref{s4-1} ($\tau_0 \sim 10^{-1}$--$10^3$), the cumulative 
effects of these neglected transitions may become non-negligible over repeated scatterings.
However, the inclusion of the transitions among excited levels would require significant 
changes to the formalism of the present work and is left for the future.

Fourth, satellite lines from lower ionisation stages are not included. These lines often
blend with the He-like K-series lines \citep{G1972} and cause the resonant Auger
destruction (RAD), in which $x$ and $y$ line photons are resonantly absorbed through the
$\mathrm{1s^2\,2s}\rightarrow\mathrm{1s\,2s\,2p}$ transition and subsequently destroyed
by Auger decay \citep{M2015,C2020b}. This process will be addressed in our next
implementation of Li-like and lower ionisation stages.

Finally, additional high-density effects, such as continuum lowering, suppression of
dielectronic recombination, three-body recombination, and free-free heating
\citep{K2021}, are not included in the present implementation. These processes can
modify the ionisation balance, recombination rates, and thermal structure at
sufficiently high densities; for example, dielectronic recombination can be
substantially suppressed at $n_{\rm e}\sim10^{10}$--$10^{15}\ {\rm cm^{-3}}$ for some
ions, while three-body recombination becomes important at $n_{\rm e}\gtrsim10^{18}\ {\rm
cm^{-3}}$, and continuum lowering and free-free heating also become increasingly
important at high densities depending on ion charge and radiation field \citep{K2021}.

\section{Summary}\label{s7}
We have presented the implementation, verification, and demonstration of the He-like
K-series lines in the MCRT code \texttt{SKIRT}. This work represents another step
towards a comprehensive MCRT treatment of ionic processes. Our main results are as
follows:

\begin{enumerate}
 \item We implemented the atomic processes responsible for the formation of the He-like
 K-series lines---resonance scattering, radiative recombination, and inner-shell
 ionisation of Li-like ions---which contribute differently to the He$\alpha$ $w$, $x$,
 $y$, and $z$ lines.
 \item The implementation successfully reproduces key RT effects,
 including the distinct line-profile evolution of the He$\alpha$ lines with optical
 depth and the formation of P Cygni profiles in outflowing gas. The results agree well
 with analytical solutions and the independent RT code \texttt{Cloudy}.
 \item We demonstrated the treatment of linear polarisation associated with resonance
 scattering, successfully reproducing the laboratory experiment of
 \citet{R2013}.
 \item These RT effects are now observable with XRISM but cannot be
 modelled adequately by conventional 1D RT codes, highlighting the importance of 3D
 MCRT calculations with ionic processes implemented in \texttt{SKIRT}.
\end{enumerate}

\begin{acknowledgements}
 We thank the anonymous reviewer for providing useful comments for improvement.
N.\,S. acknowledges the hospitality of colleagues at Ghent University, ESTEC, and SRON,
where he stayed during the course of this work. This research made use of the JAXA
high-performance computing system JSS3. N.\,S. acknowledges support from the JSPS
Core-to-Core Program (grant number JPJSCCA20220002), the Foundation for Promotion of
Astronomy, and JSPS KAKENHI Grant Number 26H02075. B.\,V. acknowledges support through
the European Space Agency (ESA) Research Fellowship Programme in Space Science.
\end{acknowledgements}

\bibliographystyle{aa}
\bibliography{main}
\end{document}